\documentclass[a4paper,11pt]{article}
\pdfoutput=1 

\usepackage{jcappub} 

\usepackage[T1]{fontenc} 

\usepackage{ amssymb }

\usepackage{soul}
\usepackage{bm}
\usepackage[utf8]{inputenc}

\title{\boldmath  Dynamical Solution to the Eta Problem in Spectator Field Models}

\author[a,b,1]{Sana Elgamal,\note{Corresponding author.}}
\author[a,b,c]{Keisuke Harigaya}

\affiliation[a]{Department of Physics, University of Chicago, Chicago, IL 60637, USA}
\affiliation[b]{Kavli Institute for Cosmological Physics and Enrico Fermi Institute,
University of Chicago, Chicago, IL 60637, USA}
\affiliation[c]{Kavli Institute for the Physics and Mathematics of the universe (WPI),
The University of Tokyo Institutes for Advanced Study,
The University of Tokyo, Kashiwa, Chiba 277-8583, Japan}

\emailAdd{selgamal@uchicago.edu}
\emailAdd{kharigaya@uchicago.edu}

\abstract{We study a class of spectator field models that addresses the eta problem while providing a natural explanation for the observed slight deviation of the spectrum of curvature perturbations from scale-invariance.
In particular, we analyze the effects of quantum corrections on the quadratic potential of the spectator field given by its gravitational coupling to the Ricci scalar and the inflaton energy, so-called the Hubble-induced mass term. 
These quantum corrections create a minimum around which the potential is flatter and to which the spectator field is attracted. We demonstrate that this attractor dynamics can naturally generate the observed slightly red-tilted spectrum of curvature perturbations. Furthermore, focusing on a curvaton model with a quadratic vacuum potential, we compute the primordial non-Gaussianity parameter $f_{\text{NL}}$ and derive a predictive relationship between $f_{\text{NL}}$ and the running of the scalar spectral index. This relationship serves as a testable signature of the model. Finally, we extend the idea to a broader class of models where the spectator field is an angular component of a complex scalar field.  
}

\begin{document}
\maketitle
\flushbottom

\section{Introduction}
\label{sec:intro}
The origin of structure in the universe is not yet known and is one of the most important topics in cosmology.
The standard scenario posits that a scalar field called the inflaton is responsible for driving inflation as well as the generation of curvature perturbations. A priori, however, this need not necessarily be the case. Any light scalar field acquires fluctuations in its field value during inflation~\cite{Mukhanov:1981xt,Hawking:1982cz,Starobinsky:1982ee,Guth:1982ec,Bardeen:1983qw}, thereby contributing to the generation of cosmic perturbations. This possibility is particularly plausible in physics beyond the Standard Model where there are numerous other light scalar fields besides the inflaton, such as in supersymmetric models. These models where the curvature perturbations are sourced by a light scalar field that has negligible influence on the inflationary expansion are known as spectator field models.
Spectator field models include the curvaton~\cite{Mollerach:1989hu, Linde:1996gt, Enqvist:2001zp, Lyth:2001nq, Moroi:2001ct} and the modulated reheating model~\cite{Dvali:2003em, Kofman:2003nx}.
In this work, we consider generic spectator field models and denote the spectator field by $\sigma$.

In spectator field scenarios, the inflaton field no longer generates the dominant part of the curvature perturbations, thereby significantly relaxing constraints on inflation~\cite{Moroi:2005kz}. For instance,
two simple classes of models called chaotic inflation~\cite{Linde:1983gd} and natural inflation~\cite{Freese:1990rb} do not necessarily predict large tensor fluctuations and hence become viable in this framework. Furthermore, the eta problem of the inflaton~\cite{Ovrut:1983my,Holman:1984yj,Goncharov:1983mw,Coughlan:1984yk,Copeland:1994vg} is alleviated:~the second slow-roll parameter of the inflaton does not have to be as small as $O(0.01)$ to explain the nearly scale-invariant spectrum of cosmic perturbations, and it is only required that inflation lasts long enough. For small-field inflation models with a small first slow-roll parameter, even values of $\eta = O(0.1-1)$ are allowed; see Appendix~\ref{app:eta} for further discussion.
Spectator field models also have interesting observational signatures. In particular,
they generically predict a non-Gaussianity of cosmic perturbations that is significantly larger than those expected in single-field inflation models, which may be discovered by near-future observations.

The alleviation of the eta problem is particularly beneficial for the landscape scenario, where the inflaton potential should be least fine-tuned while satisfying the anthropic requirements~\cite{Vilenkin:1994ua,Tegmark:2004qd,Freivogel:2005vv}.
In particular, the eta parameter of the inflaton is not naturally small, since  inflation can occur even with $|\eta| > 0.01$. This makes the observed nearly scale-invariant spectral index ($n_{s} \simeq 0.96$~\citep{Planck:2018vyg}) difficult to achieve without  unnecessary fine-tuning.
Spectator field models can address this challenge by decoupling the generation of perturbations from the inflaton.

Nevertheless, spectator field models generically suffer from their own eta problem. The eta parameter of the spectator field is $\eta_\sigma \equiv M_{\rm pl}^{2}\frac{V''}{V_{\rm inf}}$, where primes denote derivatives with respect to $\sigma$, $V_{\rm inf}$ is the inflaton potential energy, and $M_{\rm pl}$ is the reduced Planck mass. The spectral index of curvature perturbations is given by
\begin{equation}
    n_s = 1- 2 \epsilon + 2 \eta_\sigma,
\end{equation}
where $\epsilon$ is the first slow-roll parameter of the inflaton. In order to explain the observed spectral index $n_s\simeq 0.96$, $|\eta_\sigma| \lesssim O(0.01)$ is required. However, $\sigma$ generically has gravitational couplings to the Ricci scalar $R$ and the inflaton potential, which impart to $\sigma$ a mass squared of $O(H^2)$, commonly referred to as the Hubble-induced mass. This Hubble-induced mass leads to $\eta_\sigma = O(1)$. Without an additional mechanism to suppress this contribution, the eta problem persists in spectator field models.

Although a small $\eta_\sigma$ may be simply attributed to a small coupling, it would be more compelling to instead have some mechanism that dynamically explains the smallness of $\eta_{\sigma}$ as well as the observed slight deviation from scale-invariance. In this paper, we demonstrate that quantum corrections to the Hubble-induced mass term provide such a mechanism. These corrections dynamically relax $\eta_\sigma$ to be a typical loop factor of $O(0.01)$, making the observed slight deviation from scale-invariance $|n_s-1| =O(0.01)$ a {\it prediction} of the theory.

This paper is structured as follows. In Sec.~\ref{sec:curvaton}, we discuss the quantum corrections to the Hubble-induced mass and show that $\eta_\sigma$ is $O(0.01)$ around the minimum of the potential to which $\sigma$ is attracted. We then derive the evolution of the spectral index for our spectator field potential during inflation, evaluate the naturalness of the observed spectral index by defining a fine-tuning measure, and compute the running of the spectral index $\alpha_s$.
Focusing on a curvaton model with a quadratic vacuum potential, 
we compute the non-Gaussianity parameter $f_{\text{NL}}^{\text{local}}$
and derive a correlation between $f_{\text{NL}}^{\text{local}}$ and $\alpha_{s}$.
In Sec.~\ref{sec:axion_curvaton}, we apply the same idea to a different class of models where the
spectator field is the angular component of a complex scalar field. We summarize our findings in Sec.~\ref{sec:conc}.

\section{Spectator field with running mass}
\label{sec:curvaton}
In this section, we examine the potential of the spectator field with quantum corrections taken into account. The quantum corrections introduce the running of the Hubble-induced mass, i.e., a logarithmic dependence of the mass on the spectator field value, flattening the potential around the minimum. Additionally, we analyze the evolution of the spectator field, the spectral index and its running, and non-Gaussianity. Our computation is generic in the sense that it does not rely on any specific form of the inflaton potential. 
With the exception of Sec.~\ref{sec:NG}, the analysis presented in this section is applicable to generic spectator field models where
the curvature perturbations $\zeta \propto \delta \sigma$, and ${\cal P}_{\delta\sigma}(k)$ is determined by the potential of $\sigma$. 

\subsection{Potential of the spectator field}

The Lagrangian ${\cal L}_\sigma$ of the spectator field is given by
\begin{equation}
\label{eq:lagrangian}
\frac{1}{\sqrt{-g}}\mathcal{L}_{\sigma} = \frac{1}{2}{\dot{\sigma}}^{2} - \frac{1}{2a^2}(\nabla{\sigma})^{2}-V(\sigma),
\end{equation}
where
$g$ is the determinant of the metric tensor, dots denote derivatives with respect to cosmic time, and $a$ is the scale factor of the universe.  
The potential of the spectator field $V(\sigma)$ is made up of two components. The first part, the vacuum potential $V_{\rm vac}(\sigma)$, depends solely on the spectator field itself, and is given by
\begin{equation}
\label{eq:vac_potential}
V_{\rm vac}(\sigma) = \frac{1}{2}m_{\sigma}^{2}\sigma^{2} + \cdots,
\end{equation}
where $m_{\sigma}$ is the mass of the spectator field and the dots in Eq.~\eqref{eq:vac_potential} denote higher order terms. The second part of the potential, namely $V_H(\sigma)$, arises due to the gravitational couplings of $\sigma$ with the Ricci scalar $R$ and the inflaton potential $V_{\rm inf}$,
\begin{align}
    V_H(\sigma) \simeq c_1 R \sigma^2 + c_2 \frac{V_{\rm inf}}{M_{\rm pl}^2} \sigma^2 \equiv c H^2\sigma^2,
\end{align}
 where $c_1$, $c_2$, and $c$ are $O(1)$ constants and $H \equiv \frac{\dot{a}}{a}$ is the Hubble parameter. In the second equality, we used $R\sim V_{\rm inf}/M_{\rm pl}^2\sim H^2$, valid when the inflaton dominates the universe.
This Hubble-dependent potential serves as an effective mass term for $\sigma$, and is commonly referred to as the Hubble-induced mass.

We now turn to the spectral index $n_{s}$, which is given by \citep{Lyth:2009zz}
\begin{equation}
\label{eq:spec_index}
n_{s} - 1 = 2 \frac{\dot{H}_{\ast}}{H_{\ast}^{2}} + \frac{2}{3} \frac{V''_{\ast}}{H_{\ast}^{2}},
\end{equation}
where throughout the asterisks denote that the variables are evaluated
at the horizon exit during inflation.
Assuming that inflation is driven by a canonical slow-rolling inflaton, the first term can be written as
\begin{equation}
\label{eq:Hdot}
2\frac{\dot{H}_{\ast}}{H_{\ast}^{2}} = -2\epsilon = -\frac{1}{M_{\rm pl}^2} \left(\frac{\mathrm{d}\phi}{\mathrm{d}N}\right)^{2},
\end{equation}
where $\phi$ is the inflaton field and $N$ is the number of inflationary e-folds. Unless the inflaton field transverses many Planck distances in field space during inflation, the first term in Eq.~\eqref{eq:spec_index} is negligibly small.  
We can therefore approximate Eq.~\eqref{eq:spec_index} as
\begin{equation}
\label{eq:spec_index_approx}
n_{s} - 1 \simeq \frac{2}{3} \frac{V''_{\ast}}{H_{\ast}^{2}}.
\end{equation}

From Eq.~\eqref{eq:spec_index_approx}, achieving a nearly scale-invariant spectrum, where $|n_{s}-1| \ll 1$, requires $|V''_{\ast}| \ll H^{2}_{\ast}$.
Barring cancellation, this implies that $|c|$ and $V_{\rm vac, \ast}''/H_{\ast}^2 \ll 1$.  
Although this small mass of the spectator field could naturally arise as a result of shift symmetry~\cite{Dimopoulos:2003az},
invoking shift symmetry restricts the possible candidates for spectator fields.
Furthermore, since the observed spectral index slightly deviates from unity, the shift symmetry must be explicitly broken to allow $V'' = O(0.01) H^2$,
which generically requires a coincidence of two unrelated mass scales, namely the Hubble scale during inflation and the spectator field mass.

In this paper, we propose an alternative scenario where $n_s\simeq 0.96$ naturally arises and that can be applied to a broader class of models. Generically, the Hubble-induced mass term can receive quantum corrections due to quartic interactions between $\sigma$ and other scalar fields with Hubble-induced masses. These corrections
introduce the renormalization group running of the Hubble-induced mass, namely, a logarithmic dependence of the mass on $\sigma$,
\begin{equation}
\label{eq:VH_corrected}
V_H(\sigma) = cH^{2} \sigma^{2} + bH^{2} \sigma^{2}\ln\left({\frac{\sigma}{M}}\right) \equiv b H^2 \sigma^2 \left[{\ln}\left(\frac{\sigma}{\sigma_0}\right)- \frac{1}{2}\right],
\end{equation}
where $M$ is the UV energy scale at which the Hubble-induced mass is set, and $b$ is given by the product of the loop factor $O(\frac{1}{8\pi^2})$, the quartic coupling constant, and the number of particles that couple to $\sigma$.
In the second equality, the potential is re-expressed in terms of the field value $\sigma_0$ defined to be that at the extremum of $V_H$ ($V_H'(\sigma_0)\equiv 0$),

\begin{equation}
\label{eq:sigma0}
\sigma_{0} = M \exp{\left(-\frac{1}{2} -\frac{c}{b}\right)}.
\end{equation}
Assuming $b>0$, $\sigma_{0}$ corresponds to the minimum of the potential. In particular, the quantum corrections create a minimum around which the Hubble-induced mass of
$\sigma$ is effectively suppressed by $b = O(0.01-0.1)$,
assuming that the coupling constants are $O(1)$. This modification of the potential by the running is analogous to the Peccei-Quinn symmetry breaking by the running of a soft mass in supersymmetric theories~\cite{Moxhay:1984am}.

\begin{figure*}
\centering
\includegraphics[width=300pt, trim={0.16cm 0.11cm 0.11cm 0.11cm}, clip] {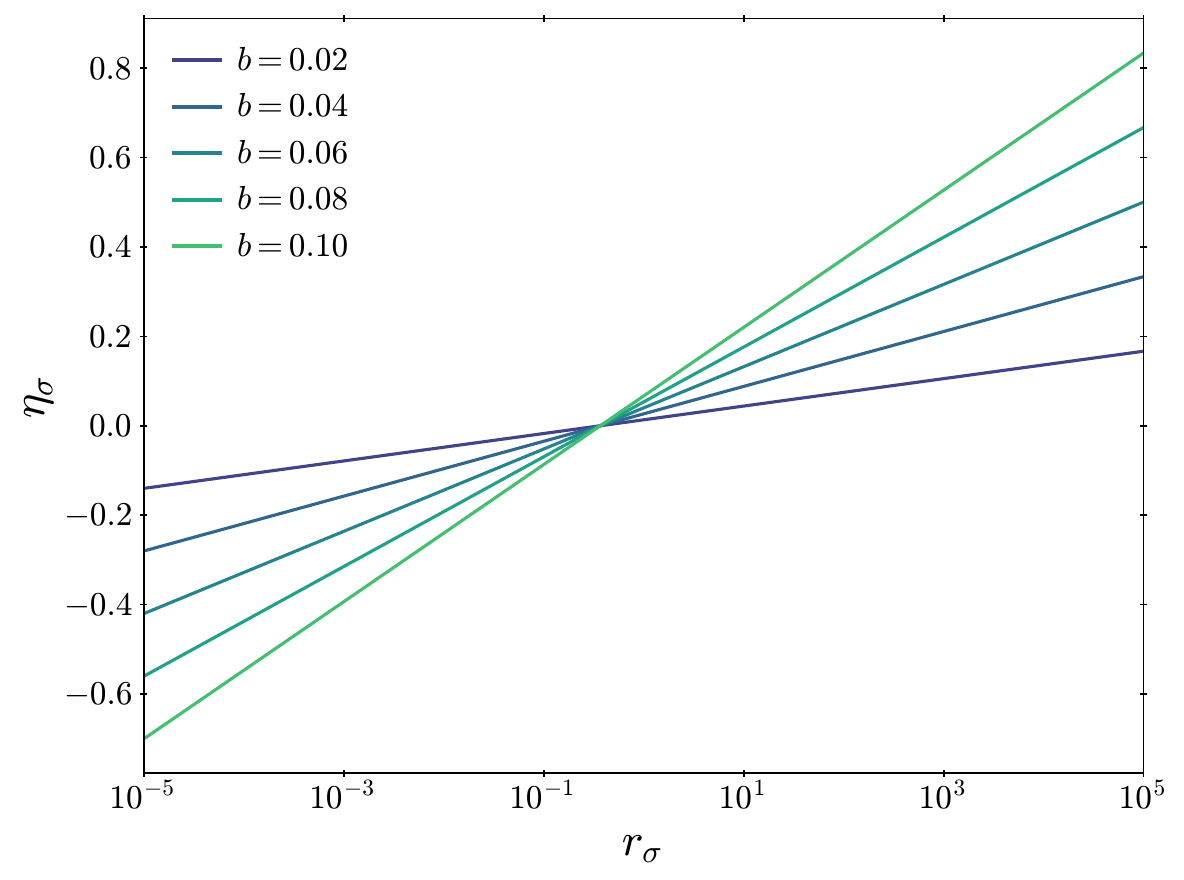}
\caption{The eta parameter of the spectator field as a function of $r_{\sigma} = \sigma/\sigma_0$ for different values of $b$, where $b \in [0.02, 0.1]$. The quantum corrections create a minimum $(r=1)$ around which the eta parameter is $O(0.01)$. 
}
\label{fig:curv_potential}
\end{figure*}

It is convenient to parametrize the potential in Eq.~\eqref{eq:VH_corrected} in the following form,
\begin{equation}
\label{eq:rep_curv_potential}
V_{H}(r) =  br_{\sigma}^{2} \left(\ln{r_{\sigma}} - \frac{1}{2} \right)\times H^{2} \sigma_{0}^{2},
\end{equation}
where $r_{\sigma} \equiv \frac{\sigma}{\sigma_{0}}$ is the field value normalized to the minimum $\sigma_{0}$. Fig.~\ref{fig:curv_potential} shows the behavior of the eta parameter for $\sigma$ across different values of $b$.
The results demonstrate that the quantum corrections drive $\eta_{\sigma}$ to $O(0.01)$ near the minimum, solving the eta problem. As we shall see in the next subsection, achieving the observed spectral index requires $r_{\sigma} < 1$ when the CMB scale exits the horizon during inflation. In fact, this condition can be naturally satisfied if the initial field value is $\sigma_{i} \simeq 0$. Since $\sigma=0$ is a symmetry enhanced point, this initial condition can be explained by a positive mass for $\sigma$ prior to observable inflation. Such a mass could arise either from a positive Hubble-induced mass around the origin or from a positive thermal mass.

The interactions responsible for the quantum corrections to the Hubble-induced mass term also generically renormalizes the self-quartic coupling of the spectator field. Although such corrections would ruin the flatness of the spectator field potential, these are suppressed in supersymmetric theories because of the non-renormalization of the superpotential~\cite{Grisaru:1979wc,Seiberg:1993vc}. Consequently, our scenario is most effectively realized within supersymmetric theories. Supersymmetric theories also predict many light scalar fields, making it plausible for one of these fields to serve as the spectator field. We note that the renormalization to the self-quartic coupling in non-supersymmetric theories has been studied in the literature as a means to flatten the inflaton potential in chaotic inflation models~\cite{Senoguz:2008nok,Enqvist:2013eua,Ballesteros:2015noa}, although quadratically divergent corrections to the mass of the inflaton are generically not under control without supersymmetry.

The flattening of the scalar potential by quantum corrections to the Hubble-induced mass term is noted in~\cite{Stewart:1996ey} for an inflaton potential. In fact, taking $\sigma$ to be the inflaton field in Eq.~\eqref{eq:VH_corrected}, one may add a (nearly) constant potential term to drive inflation. However, such a setup generically suffers from fine-tuning problems. For $b>0$, the potential becomes flat around the minimum of the potential, so inflation cannot end unless one introduces a waterfall field that couples to $\sigma$ such that  the waterfall field begins rolling around $\sigma = \sigma_0$. Since $\sigma=\sigma_0$ is not a symmetry enhanced point, this requires the fine-tuning of parameters. On the other hand, for $b<0$, the potential flattens around the maximum of the potential. For $\sigma < \sigma_0$, the inflaton field rolls towards $\sigma =0$, and if a waterfall field is coupled to $\sigma$, the waterfalling at $\sigma \ll \sigma_0$ can end the inflation. However, this setup has an initial condition problem, as it must explain why the inflaton field is set around the maximum of the potential. Since the maximum of the potential is not a symmetry enhanced point, it is not possible to set the initial condition at the maximum through a positive thermal or Hubble-induced mass term. In~\cite{Stewart:1997wg}, it is suggested that the initial condition may be set by tunneling from $\sigma \gg \sigma_0$ to $\sigma \lesssim \sigma_0$ whose rate is suppressed. For $\sigma > \sigma_0$, the inflaton rolls toward larger field values, and the slow-roll condition may be violated around $\sigma \sim M_{\rm pl}$, ending the inflation. Nevertheless, the initial condition around $\sigma_0$ still requires an explanation.
In contrast, in our setup, the required initial condition of $\sigma$ is around the origin, which can be naturally obtained as explained above.

\subsection{Evolution of the spectral index}

We now express the spectral index as a function of the model parameters.
Using the expression for the potential given by Eq.~\eqref{eq:VH_corrected}, the spectral index is
\begin{equation}
\label{eq:n_b}
n_{s}(N) \simeq 1 + \frac{4}{3}b(1+\ln\{r_{\sigma}(N)\}),
\end{equation}
where $N$ is the number of e-folds during inflation and serves as the time variable in our analysis. We adopt the forward-counting convention in which $N$ increases as inflation progresses, with $N=0$ corresponding to the time when the field begins rolling from its initial value. Consequently, 
$N=0$ does not necessarily coincide with the moment when the CMB pivot scale exits the horizon.

Using the equation of motion of $\sigma$ in the slow-roll approximation,
\begin{equation}
\label{eq:slowroll}
3H\dot{\sigma} + V'\simeq 0,
\end{equation}
and the parametrization of the potential in Eq.~\eqref{eq:rep_curv_potential}, the equation of motion of $r_{\sigma}(N)$ becomes
\begin{equation}
\label{eq:KG_slow_roll}
3\frac{\mathrm{d}r_{\sigma}}{\mathrm{d}N} +2 b r_{\sigma}\ln{r_{\sigma}} \simeq 0,
\end{equation}
whose solution is
\begin{equation}
\label{eq:rsol_inf}
{\ln}\{r_{\sigma}(N)\} = {\rm exp}\left(- \frac{2b N}{3}\right){\ln}(r_{\sigma,i}),
\end{equation}
where $r_{\sigma,i} \equiv r_{\sigma}(0)$ is the initial normalized field value. Plugging this into Eq.~\eqref{eq:n_b}, we find
\begin{equation}
\label{eq:n_s_analytical}
n_{s}(N) \simeq 1 + \frac{4}{3}b \left[1+ \exp\left(-\frac{2}{3}bN \right)\ln{(r_{\sigma,i})} \right].
\end{equation}

\begin{figure*}
\centering
\includegraphics[width=450pt, trim={0.16cm 0.11cm 0.11cm 0.11cm}, clip] 
{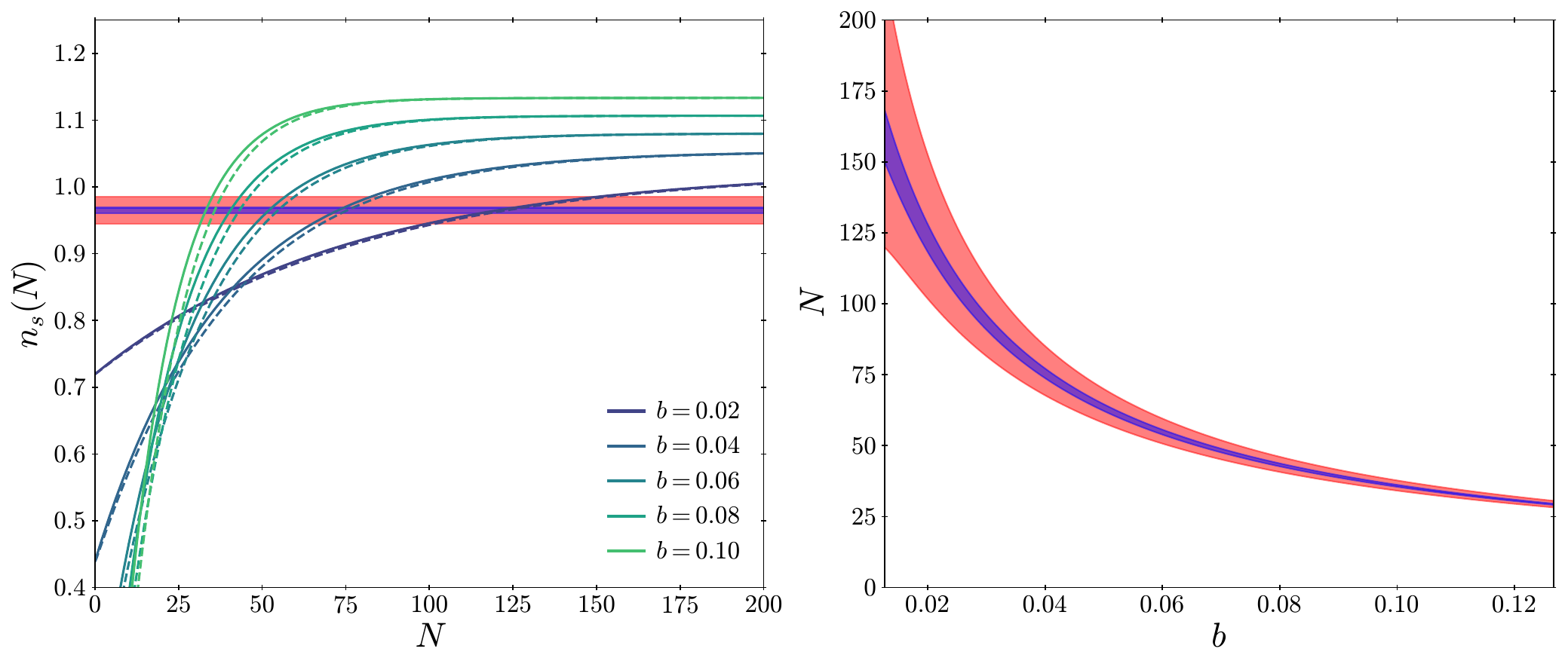}
\caption{
The left panel shows the spectral index verses the number of inflationary e-folds for $b \in [0.02,0.1]$ and fixed $r_{\sigma,i} = 10^{-5}$. The solid curves show the spectral index computed 
using the slow-roll approximation as in Eq.~\eqref{eq:n_s_analytical}, while the dashed curves show the spectral index evaluated by numerically computing $r_{\sigma}(N)$. The blue-shaded region highlights areas where the  scalar spectral index is $n_{s} = 0.965 \pm 0.004$ as measured by \textit{Planck} at the $1\sigma$ level~\citep{Planck:2018vyg}, while the red-shaded region corresponds to areas where $|0.965-n_{s}| \le 0.02$.
The right panel shows the range of the number of inflationary e-folds as a function of $b$, where the shaded regions correspond to the same ranges of $n_{s}$ as those in the left panel.}
\label{fig:specindexvsN1}
\end{figure*}

The left panel of Fig.~\ref{fig:specindexvsN1} shows the spectral index as a function of the number of inflationary e-folds $N$ for different values of $b$.
While the classical initial condition set by the positive Hubble-induced mass or thermal mass is $\sigma_i=0$, we choose $r_{\sigma,i}=10^{-5}$ since quantum fluctuations introduce deviations of $\sigma_i\sim H \sim {\cal P}_\zeta^{1/2} \sigma_0\simeq 10^{-5} \sigma_0$, where we used $\zeta \sim \delta \sigma /\sigma$. The solid and dashed curves show $n_s$ where $r_{\sigma}(N)$ is computed analytically as in Eq.~\eqref{eq:rsol_inf} using the slow-roll approximation and numerically without the approximation, respectively. Based on the agreement of the two curves, we adopt the slow-roll approximation in the following.
The blue-shaded region
marks areas where $n_s=0.965 \pm 0.004$ as measured by \textit{Planck} at the $68\%$ confidence level.

The red-shaded region
highlights the extended range $n_s=0.965 \pm 0.02$, which can be considered to be qualitatively similar to the observed one, since it describes a slightly red-tilted spectrum (i.e., $1-n_{s} = O(0.01)$). This range can be used to qualitatively assess the naturalness of the model for a given value of $b$. We emphasize that the choice of $n_s=0.965 \pm 0.02$ as a measure of typicality is intended to capture the qualitative feature that the observed spectral index lies below unity by about 0.04, rather than to impose the precise experimental error bars.\footnote{Using the observational window would make the notion of typicality unstable, since the allowed range of e-folds would shrink as measurements become more precise, artificially suggesting increasing fine-tuning. In contrast, the broader criterion that we use provides a stable measure of how long the spectral index remains qualitatively similar to its observed value during inflation.}

The right panel of Fig.~\ref{fig:specindexvsN1} shows the range of number of e-folds corresponding to the shaded regions in the left panel. For smaller values of $b$, the field $\sigma$ stays within these regions for a longer duration, indicating that the observed spectral index is likely a typical outcome of this model. One can see that for $b=0.01-0.1$, the CMB scale should exit the horizon when $ N\approx20-150$ after $\sigma$ begins to roll from the origin. Typically, the number of e-folds after the CMB scale exits the horizon is $50-60$. This means that the last inflation~---~the inflation accounting for the observed cosmic perturbations~---~should span a maximum total number of e-folds of $\approx 200$. This, however, does not preclude the possibility of a longer total inflationary period, including an eternal one preceding 
the last inflation, provided that the spectator field remains trapped at the origin before the onset of the last inflationary phase.

\begin{figure*}
\centering
\includegraphics[width=250pt, trim={0.16cm 0.11cm 0.11cm 0.11cm}, clip] 
{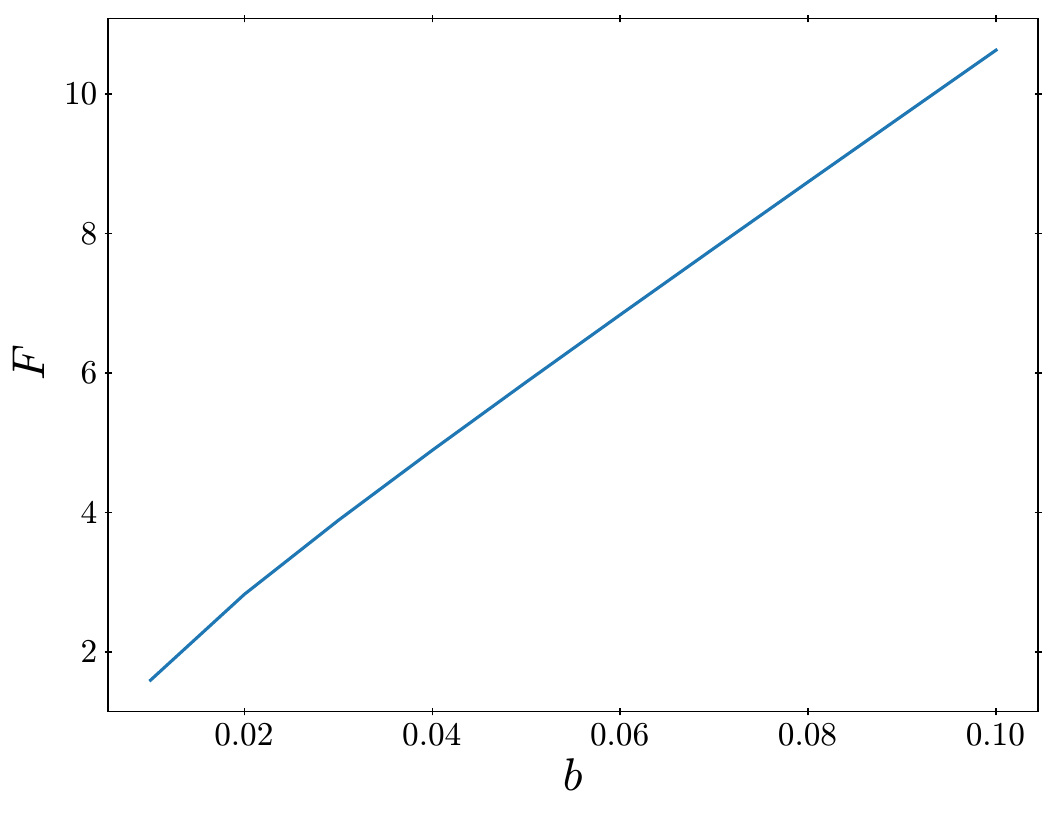}
\caption{The fine-tuning measure $F$ as a function of our model parameter $b$. The results show that when $b < 0.02$, $F<3$ meaning that the required fine-tuning is minimal. On the other hand, when $b > 0.1$, $F>10$, indicating that our model becomes increasingly unnatural for larger values of $b$.}
\label{fig:ns_naturalness}
\end{figure*}

We quantify the naturalness of the observed spectral index using the following fine-tuning measure introduced in~\cite{Barbieri:1987fn} for the electroweak scale,
\begin{align}
F \equiv {\rm max}\left(|F_{r_i}|, |F_{b}|, |F_{N}|\right),~~~
    F_X \equiv \frac{1}{n_s-1} \frac{\partial (n_s-1)}{\partial {\ln} X},
\end{align}
where $F_{X}$ quantifies the sensitivity of the observable $n_{s}-1$ to variations in a given model parameter $X$. A large $F_{X}$ indicates that small changes in $X$ lead to large changes in $n_{s}-1$, implying a high degree of fine-tuning. The overall fine-tuning measure $F$ is defined as the maximum sensitivity across all model parameters under consideration. 
Fig.~\ref{fig:ns_naturalness} presents the dependence of this fine-tuning measure $F$ on the model parameter $b$. For each value of $b$, the parameter $N$ is chosen so that $n_s=0.965$, with the initial field value set to $r_{\sigma,i} = 10^{-5}$. As shown in Fig.~\ref{fig:ns_naturalness}, when $b > 0.1$, $F>10$, implying that more than $10\%$ fine-tuning is required to obtain the observed spectral index. In contrast, for $b<0.02$, the required fine-tuning is negligible ($F<3$), suggesting that the model operates naturally in this regime.
However, it is important to note that $b<0.02$ requires an extended duration of the last inflation as shown in Fig.~\ref{fig:specindexvsN1}, which, depending on the specific inflation model, may require fine-tuning of the inflaton potential. The naturalness of the model should ultimately be assessed within the broader context of the entire inflationary scenario.

\subsection{Running of the spectral index}
One can also compute the running of the spectral index $\alpha_{s}$, which is defined as
\begin{equation}
\label{eq:n_running}
\alpha_{s} \equiv \frac{\mathrm{d}n_{s}}{\mathrm{d}\ln{k}} = \frac{\mathrm{d}n_{s}}{\mathrm{d}N} \frac{\mathrm{d}N}{\mathrm{d}\ln{k}} = \frac{1}{1-\epsilon}\frac{{\mathrm{d}}n_{s}}{{\mathrm{d}}N} \simeq \frac{{\mathrm{d}}n_{s}}{{\mathrm{d}}N},
\end{equation}
where $k$ is the comoving wavenumber of a primordial mode and we have used that $\epsilon \ll 1$ in the last equality. Using Eq.~\eqref{eq:KG_slow_roll} yields
\begin{equation}
\label{eq:running}
\alpha_{s} = - \frac{8}{9}b^{2}\ln\{r_{\sigma}(N)\}.
\end{equation}
Using Eq.~\eqref{eq:n_b}, we obtain
\begin{equation}
\label{eq:running_ns_slowroll}
\alpha_{s} = \frac{2}{3}b(1-n_{s}) + \frac{8}{9}b^{2}.
\end{equation}

\begin{figure*}
\centering
\includegraphics[width=270pt, trim={0.16cm 0.11cm 0.11cm 0.11cm}, clip] 
{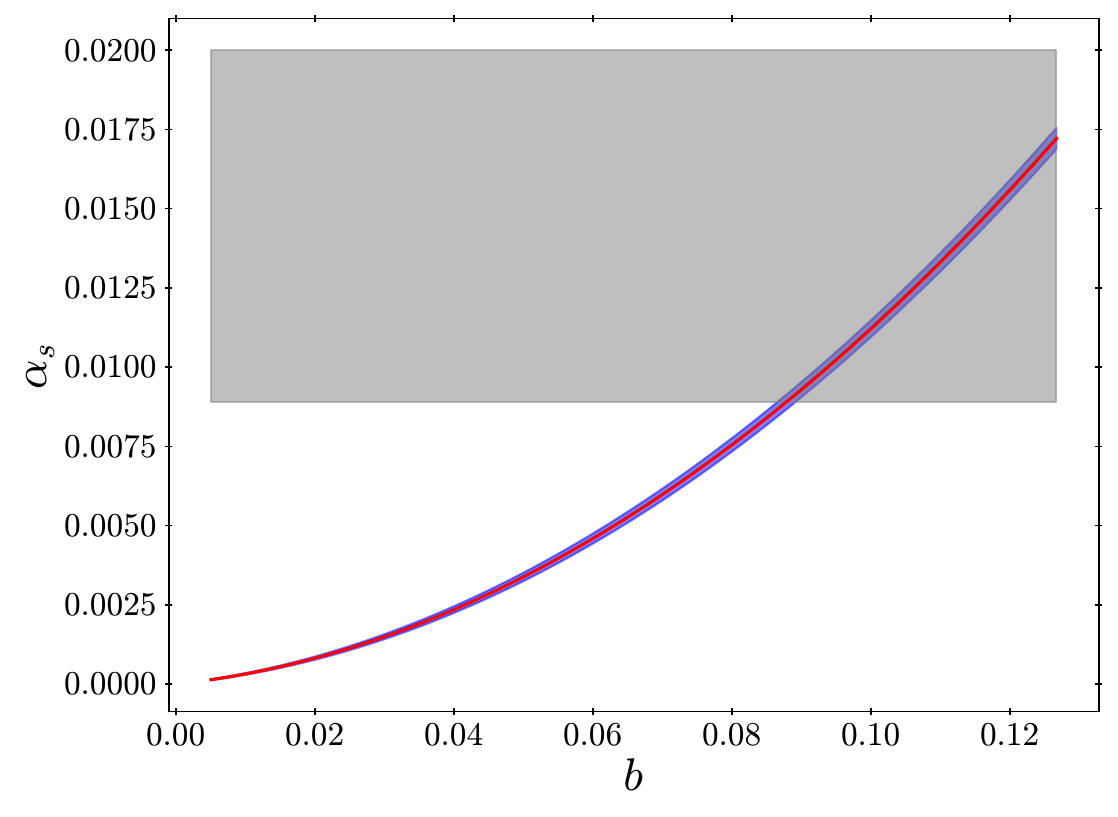}
\caption{
The red curve shows the running of the spectral index at $n_{s} = 0.965$ as a function of the potential parameter $b$. The blue-shaded region represents $n_{s}=0.965\pm0.004$, consistent with the $1\sigma$ constraints from \textit{Planck} measurements. The gray-shaded  region is excluded by \textit{Planck} measurements of $\alpha_{s}$ at the $2\sigma$ level.}
\label{fig:running_vs_b1}
\end{figure*}

Fig.~\ref{fig:running_vs_b1} shows the running computed at $n_{s} = 0.965 \pm 0.004$ as a function of the parameter $b$. The red curve shows the central value and the blue-shaded region shows the uncertainty of $n_s$. The gray-shaded region in Fig.~\ref{fig:running_vs_b1} is excluded by
the \textit{Planck} combined temperature and polarization data~\citep{Planck:2018jri},
\begin{equation}
\label{eq:running_constraint}
\alpha_{s} = -0.0045 \pm 0.0134
\end{equation}
at the $2 \sigma$ level, where we used the $1\sigma$ constraint from \textit{Planck} and assumed that the uncertainty of $\alpha_{s}$ is that of a Gaussian distribution. 
Future observations are expected to measure $\alpha_s$ with a precision of $\Delta \alpha_s\sim 10^{-3}$~\cite{Kohri:2013mxa,SPHEREx:2014bgr,Li:2018epc}.
From Figs.~\ref{fig:ns_naturalness} and \ref{fig:running_vs_b1}, one can see that our model can naturally explain the observed $n_s$ while producing an observable $\alpha_s$\footnote{Our model predicts a positive running of the spectral index which might explain the recent tension between ACT~\cite{ACT:2025fju} and Planck~\cite{Planck:2018vyg}.}.

\subsection{Non-Gaussianity}
\label{sec:NG}

As mentioned earlier, our solution to the eta problem is applicable to generic spectator field models. In this subsection, we focus on a curvaton model with a quadratic vacuum potential, and compute the non-Gaussianity parameter as well as its correlation with the running of the spectral index. 

We use the $\delta{N}$ formalism~\cite{Sasaki:1995aw,Wands:2000dp,Lyth:2004gb}  to derive the non-Gaussianity parameter. In this formalism, the curvature perturbation $\zeta$ is computed as the difference in the number of e-folds $\delta{N}$ among different patches in the universe, with the initial slice being a fixed flat slice, and the final slice being a uniform-density slice.
Assuming that the curvaton energy density is negligible when it begins oscillation due to the vacuum potential, we can take the initial time slice to be when the curvaton begins to oscillate.
Since the curvaton decays when its decay rate is around the expansion rate (which is determined by the total energy density of the universe), the time of curvaton decay also corresponds to a uniform-density slice.

The number of e-folds between the onset of curvaton oscillation and decay is given by
\begin{equation}
\label{eq:N_oscdec}
N \equiv \ln \left(\frac{a_{\rm dec}}{a_{\rm osc}} \right),
\end{equation}
where $a_{\rm osc}$ and $a_{\rm dec}$ are the scale factors when the curvaton field starts to oscillate and decays, respectively. Since the curvaton energy density scales just like pressureless matter ($\rho_{\sigma} \propto a^{-3}$),
\begin{equation}
\label{eq:N_oscdec2}
N = \frac{1}{3} \ln \left(\frac{\rho_{\sigma_{\rm osc}}}{\rho_{\sigma_{\rm dec}}} \right) = \frac{1}{3}\ln{\left(\frac{\frac{1}{2}m_{\sigma}^{2}g(\sigma_{\ast})^{2}}{\rho_{\sigma_{\rm dec}}}\right)},
\end{equation}
where $\rho_{\sigma_{\rm osc}}$ and $\rho_{\sigma_{\rm dec}}$ are the curvaton energy densities at the onset of oscillation and just before decay, respectively, and
$g(\sigma_{\ast})$ is the curvaton field value when the oscillation begins. 
Furthermore, after the onset of curvaton oscillation, the curvaton energy density scales as $\rho_{\sigma} \propto a^{-3}$ while radiation energy density scales as $\rho_{r} \propto a^{-4}$, leading to
\begin{equation}
\label{eq:rho_scaling}
\frac{\rho_{\sigma_{\rm dec}}}{\rho_{\sigma_{\rm osc}}} = \left(\frac{\rho_{r_{\rm dec}}}{\rho_{r_{\rm osc}}}\right)^{3/4}.
\end{equation}
Assuming that radiation dominates the energy density at the onset of oscillation, $ \rho_{\rm osc} \approx \rho_{r_{\rm osc}}$,
where $\rho_{\rm osc}$ is the total energy density when the oscillation begins.
Just before decay, the total energy density is given by
$\rho_{\rm dec} = \rho_{r_{\rm dec}} + \rho_{\sigma_{\rm dec}}$. The curvaton energy density just before decay satisfies the following relationship,
\begin{equation}
\label{eq:rho_sigma_dec}
\rho_{\sigma_{\rm dec}} = \frac{1}{2}m_{\sigma}^{2}g(\sigma_{\ast})^{2}\left(\frac{\rho_{\rm dec} - \rho_{\sigma_{\rm dec}}}{\rho_{\rm osc}}\right)^{3/4}.
\end{equation}
Combining Eq.~\eqref{eq:N_oscdec2} with Eq.~\eqref{eq:rho_sigma_dec}, we find
\begin{equation}
\label{eq:N_rho}
N = \frac{1}{4}\ln{\left(\frac{\rho_{\rm osc}}{\rho_{\rm dec} - \rho_{\sigma_{\rm dec}}}\right)}.
\end{equation}

Assuming that the curvature perturbations generated by the curvaton field dominate, one can express $\zeta$ up to the quadratic term in the field perturbation $\delta \sigma$ as
\begin{equation}
\label{eq:zeta_quad}
\zeta = N' \delta \sigma + \frac{1}{2}N''(\delta \sigma)^2,
\end{equation}
where the primes denote derivatives with respect to $\sigma_{\ast}$.
The curvature perturbations can be split into a Gaussian (denoted by $\zeta_{g.}$) and a non-Gaussian part (denoted by $\zeta_{n.g.}$),
\begin{equation}
\label{eq:nongauss_form}
\zeta(\bm{x}) = \zeta_{g.} + \zeta_{n.g.} \equiv h{(\bm{x})} +  \frac{3}{5} f_{\text{NL}}^{\text{local}} h^{2}(\bm{x}) ,
\end{equation}
where $h$ is a Gaussian random field and $f_{\text{NL}}^{\text{local}}$ describes the amplitude of the non-Gaussian correction. The non-Gaussianity is of the local type, meaning that $h$ only depends on the local value of the perturbations. From Eqs.~\eqref{eq:zeta_quad} and \eqref{eq:nongauss_form}, we deduce that
\begin{equation}
\label{eq:fNL}
f_{\text{NL}}^{\text{local}} = \frac{5}{6}\frac{N''}{N'^{2}}.
\end{equation}

To find an expression for $f_{\text{NL}}^{\text{local}}$, we differentiate Eq.~\eqref{eq:N_rho} with respect to $\sigma_{\ast}$, where
$\rho_{\rm osc}$ (which is dominated by the radiation component) and $\rho_{\rm dec}$ (which is evaluated at a uniform-density slice) are independent of $\sigma_{\ast}$. Using the chain rule $\frac{\partial}{\partial{\sigma_{\ast}}} = \frac{\partial{g}}{\partial{\sigma_{\ast}}} \frac{\partial}{\partial{g}}$, we find
\begin{equation}
\label{eq:dNdsigma_star}
N' = \frac{2}{3}\frac{g'}{g}\left(\frac{\rho_{\sigma_{\rm dec}} - \frac{1}{2}g\rho_{\sigma_{\rm dec,g}}}{\rho_{\sigma_{\rm dec}}}\right),
\end{equation}
\begin{equation}
\label{eq:rho_sigma_dec_g}
\rho_{\sigma_{\rm dec},g} = \frac{m_{\sigma}^{2}g(\rho_{\rm dec} - \rho_{\sigma_{\rm dec}})}{\rho_{\rm osc}^{3/4} (\rho_{\rm dec} - \rho_{\sigma_{\rm dec}})^{1/4} + \frac{3}{8}m_{\sigma}^{2}g^{2}},
\end{equation}
where the subscript $,g$ denotes the derivative with respect to $g$. Using Eq.~\eqref{eq:rho_scaling},
\begin{equation}
\label{eq:dNdsigma_star_final}
N' = \frac{2}{3}f\frac{g'}{g},
\end{equation}
where
\begin{equation}
\label{eq:f}
f \equiv \frac{3\rho_{\sigma_{\rm dec}}}{3\rho_{\sigma_{\rm dec}} + 4\rho_{r_{\rm dec}}} \simeq \Omega_{\sigma_{ \text{dec}}},
\end{equation}
and $\Omega_{\sigma_{\text{dec}}} \equiv \frac{\rho_{\sigma_{\text{dec}}}}{\rho_{\text{dec}}}$ is the fractional curvaton energy density just before decay. If the curvaton energy density is sub-dominant, then the fluctuations in the curvaton energy density must be substantial to generate sufficiently large cosmic perturbations, which results in excessively large non-Gaussianity. Thus, in order to satisfy observational constraints coming from non-Gaussianity (see Eq.~\eqref{eq:fNL_constraint}), the curvaton must be the dominant component of the universe just before it decays. We therefore consider the limit where $\Omega_{\sigma_{\rm{dec}}}=1$. In this limit, our expression for $f_{\rm{NL}}^{\rm{local}}$ becomes~\cite{Sasaki:2006kq}
\begin{equation}
\label{eq:fNL_Om_sigma}
f_{\rm{NL}}^{\rm{local}} = \frac{5}{4}\left(\frac{g g''}{g'^{2}}-1\right). 
\end{equation}

We now compute $g$ as a function of $\sigma_*$, which is determined by the evolution of the curvaton field both during and after inflation. During inflation, the evolution is given by Eq.~\eqref{eq:rsol_inf}, and the field value at the end of inflation $\sigma_{\rm end}$ is given by
\begin{equation}
{\ln}\left(\frac{\sigma_{\rm end}}{\sigma_0}\right) = {\rm exp}\left\{- \frac{2}{3}b \left(N_{\rm end} -N_*\right)\right\}{\ln}\left(\frac{\sigma_*}{\sigma_0}\right),
\end{equation}
where $N_{\rm end}$ is the number of e-folds at the end of inflation.
After inflation, the inflaton field begins to oscillate around the minimum and behaves as matter. During this period, the Hubble-induced mass of $\sigma$ is generically different from that during inflation and is given by
\begin{equation}
\label{eq:VH_MD}
V_H(\sigma) =
- \frac{1}{2} a_{\rm m} H^2 \sigma^2 +
b_{\rm m} H^2 \sigma^2 \left[{\ln}\left(\frac{\sigma}{\sigma_{0}}\right)- \frac{1}{2}\right],
\end{equation}
where $a_{\rm m}$ and $b_{\rm m}$ are constants.
As we show in Appendix~\ref{sec:HMD}, for one-field supersymmetric inflation models, $b_{\rm m}=b$ and $a_{\rm m}=3/2$.
In more generic models, $b_{\rm m}$ and $a_{\rm m}$ may take different values.
Solving the equation of motion in Appendix~\ref{sec:Dynamics_MD}, we find
\begin{equation}
\label{eq:rsol_MD}
{\ln}\left(\frac{\sigma}{\sigma_0}\right) = {\rm exp} \left\{ -\frac{4 b_{\rm m}}{\sqrt{9 + 16 a_{\rm m}}} (N -N_{\rm end}) \right\}{\ln}\left(\frac{\sigma_{\rm end}}{\sigma_0}\right) + C'
\end{equation}
where we assumed $a_{\rm m} > -9/16$ and $C'$ is a constant independent of $\sigma_{\rm end}$.
The evolution of $\sigma$ by the Hubble-induced mass term ceases when either the vacuum potential of $\sigma$ dominates over the Hubble-induced mass, after which $\sigma$ begins to oscillate, or when reheating ends, after which the Hubble-induced mass becomes negligible. Denoting the number of e-folds at which this occurs as $N_{\rm f}$, $g(\sigma_*)$ is given by
\begin{align}
\label{eq:gsol}
{\ln}\left(\frac{g}{\sigma_0}\right) =  {\rm exp}\left\{-\frac{4 b_{\rm m}}{\sqrt{9 + 16 a_{\rm m}}} (N_{\rm f} -N_{\rm end})- \frac{2}{3}b (N_{\rm end}-N_{*}) \right\} {\ln}\left(\frac{\sigma_*}{\sigma_0}\right) +C'.
\end{align}
Using Eqs.~\eqref{eq:fNL} and \eqref{eq:gsol}, we obtain
\begin{equation}
\label{eq:fNL_2}
f_{\text{NL}}^{\text{local}} \simeq -\frac{5}{4} \exp \left \{ \frac{4 b_{\rm m}}{\sqrt{9 + 16 a_{\rm m}}} (N_{\rm f} -N_{\rm end}) + \frac{2}{3}b(N_{\rm end} - N_*) \right \} \equiv -\frac{5}{4} \exp \left\{\frac{2}{3}b\Delta N \right\},
\end{equation}
where $\Delta{N}$ parametrizes the duration of the non-harmonic curvaton evolution after horizon exit. Eq.~\eqref{eq:fNL_2} shows that $a_{\rm m}>0$ suppresses the effect of the curvaton's post-inflationary dynamics on non-Gaussianity. Additionally, a positive value of $a_{\rm m}$ enhances the curvaton field value helping the curvaton dominate the universe~\cite{Fujita:2016vfj}.

\begin{figure*}
\centering
\includegraphics[width=430pt, trim={0.16cm 0.11cm 0.11cm 0.11cm}, clip] 
{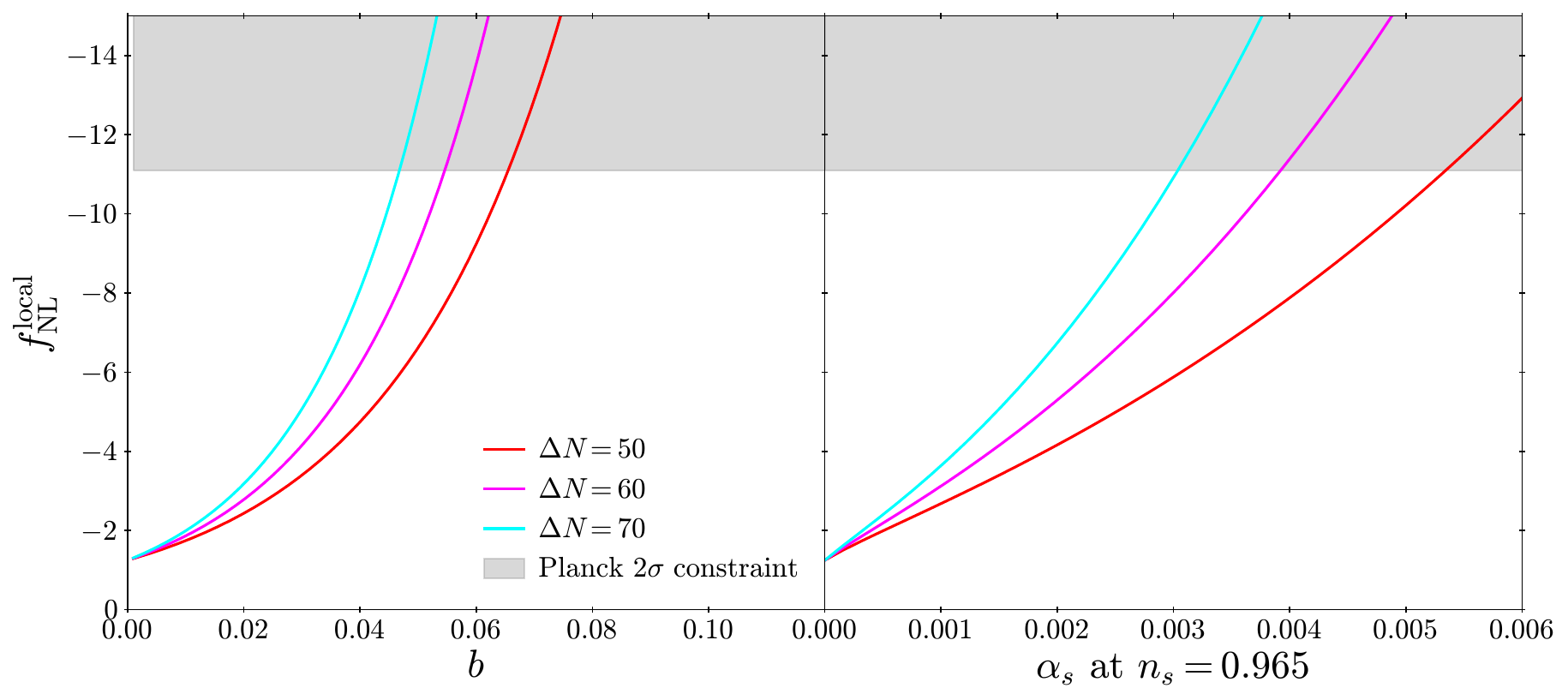}
\caption{
The non-Gaussianity parameter $f_{\text{NL}}^{\text{local}}$ as a function of $b$ for $\Delta N \in \{50,60,70\}$
assuming the curvaton dominates before its decay (left) and $\alpha_s$ at $n_{s} = 0.965$ (right), where $\Delta N$ captures the duration of the non-harmonic curvaton evolution after horizon exit as defined in Eq.~\eqref{eq:fNL_2}. The gray-shaded regions are excluded by \textit{Planck} $f_{\text{NL}}^{\text{local}}$ measurements at the $2\sigma$ level.
}
\label{fig:fNL_vs_b}
\end{figure*}

The contribution to $\Delta N$ during inflation is typically $50-60$, while post-inflationary dynamics, specifically during matter-domination, contribute an additional term
\begin{equation}
\label{eq:DN_MD}
\Delta N_{\rm MD}=\frac{4}{\sqrt{9 + 16a_{\rm m}}} \frac{b_{\rm m}}{b} {\ln}\left(\frac{H_{\rm inf}}{H_{\rm f}}\right),
\end{equation}
where $H_{\rm f}$ is the Hubble scale at $N=N_{\rm f}$, $H_{\rm inf}$ is the Hubble scale during inflation ($\lesssim 6\times 10^{13}$ GeV)%
\footnote{
If $H_{\rm inf} \gtrsim 10^{13}$ GeV, $\sigma_0$ must be around the Planck scale to explain $ P_\zeta \sim 10^{-9}$. This means that the Hubble-induced mass must be as small as $b H^{2} \ll H^{2}$ at the UV scale. To obtain the small Hubble-induced mass by its running rather than by a UV boundary condition, it is preferable that $H_{\rm inf} \ll 10^{13}$~GeV.
}, and $H_{\rm f}$ is bounded below by the soft mass scale ($\gtrsim 1$ TeV).
In supersymmetric one-field inflation models, $b_{\rm m}=b$ and $a_{\rm m}=3/2$.  For these parameters, $\Delta N_{\rm MD}$ is at most 17, and smaller values of the inflation scale or larger curvaton masses reduce this contribution further.
As a result, the total $\Delta N$ typically falls in the range $50-70$.

The left panel of Fig.~\ref{fig:fNL_vs_b} shows $f_{\text{NL}}^{\text{local}}$ as a function of $b$ for different values of $\Delta N \in \{50,60,70\}$. As $b$ increases, the curvaton potential becomes less harmonic, leading to a corresponding increase in $|f_{\text{NL}}^{\text{local}}|$. Observational constraints on local non-Gaussianity, derived from the combined temperature and polarization \textit{Planck} data, impose the following bound~\citep{Planck:2019kim}
\begin{equation}
\label{eq:fNL_constraint}
f_{\text{NL}}^{\text{local}} = -0.9 \pm 10.2
\end{equation}
at the $2 \sigma$ level, where we used the $1\sigma$ constraint from \textit{Planck} and assumed that the uncertainty is that of a Gaussian distribution. The gray-shaded region in Fig.~\ref{fig:fNL_vs_b} represents the parameter space excluded at the $2\sigma$ level.

Eqs.~\eqref{eq:running_ns_slowroll} and \eqref{eq:fNL_2} can be used to derive the following correlation between the local non-Gaussianity parameter and the running of the spectral index,
\begin{equation}
\label{eq:fNL_running_correl}
f_{\text{NL}}^{\text{local}} \simeq -\frac{5}{4}\exp \left \{\frac{\Delta{N}}{4} \left(n_{s}-1 + \sqrt{8 \alpha_{s} + (1-n_{s})^{2}}\right)\right \}.
\end{equation}
This relationship, evaluated at $n_{s} = 0.965$, is depicted in the right panel of Fig.~\ref{fig:fNL_vs_b}. Allowing for the uncertainty in the precise value of $\Delta{N}$, Eq.~\eqref{eq:fNL_running_correl} establishes a relation between $\alpha_{s}$ and $f_{\text{NL}}^{\text{local}}$ within a few ten percents. Upcoming cosmological probes targeting non-Gaussianity and the running~\cite{SPHEREx:2014bgr} offer a direct opportunity to test this prediction.

\section{Spectator field models with approximate $U(1)$ symmetry}
\label{sec:axion_curvaton}
In this section, we consider a scenario with a complex scalar field $\Sigma$, which can be decomposed into a radial component $\sigma$ and an angular component $\theta$,
\begin{equation}
\label{eq:phi_radial}
\Sigma = \frac{1}{\sqrt{2}}\sigma \exp{(i\theta)}.
\end{equation}
We assume $\Sigma$ possesses an approximate global $U(1)$ symmetry,
which guarantees that the Hubble-induced mass of $\theta$ is suppressed. As a result, if the cosmic perturbations are sourced by the fluctuations of $\theta$, the eta problem can be avoided~\cite{Dimopoulos:2003az}. However, an important question remains: why does the spectral index deviate from unity by $O(0.01)$? The scenario described below, which utilizes quantum corrections to the Hubble-induced mass of $\sigma$, can naturally explain this deviation.

\subsection{Evolution of the spectral index}

\begin{figure*}
\centering
\includegraphics[width=430pt, trim={0.16cm 0.11cm 0.11cm 0.11cm}, clip] 
{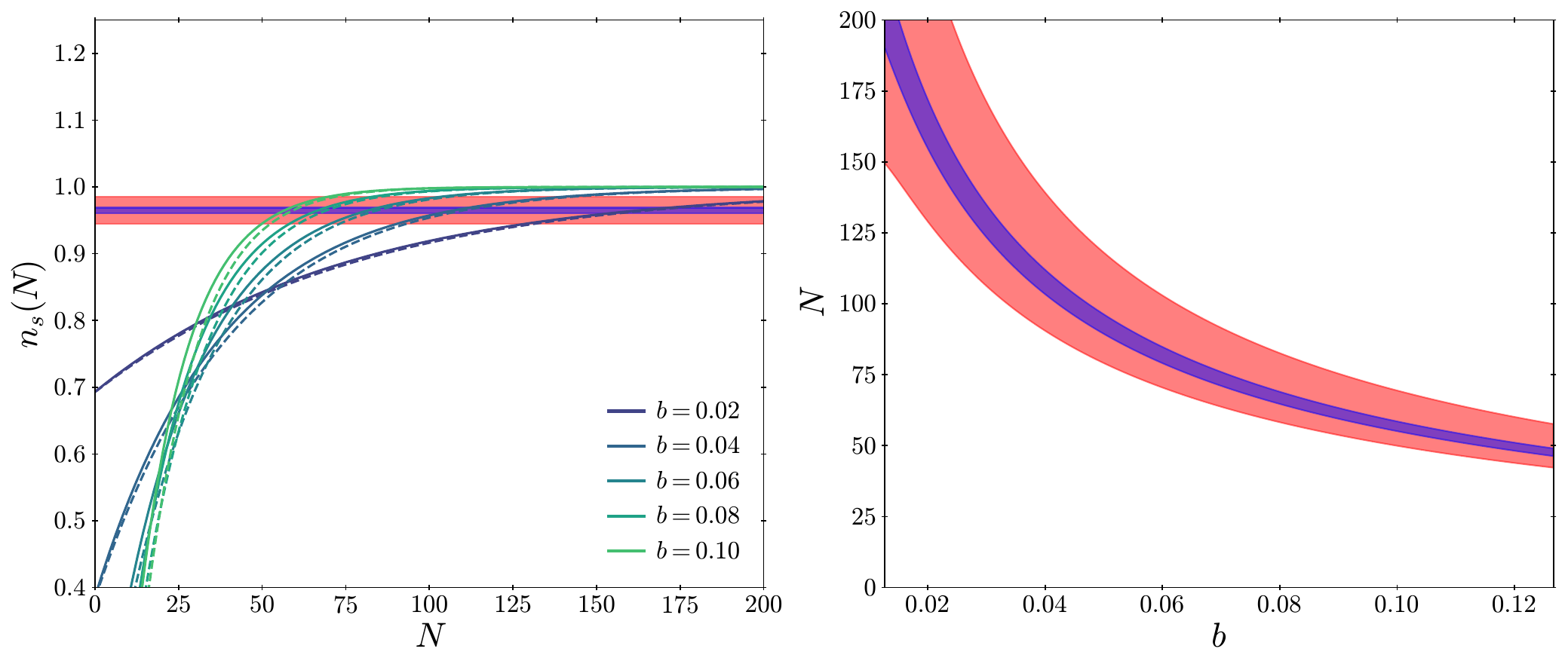}
\caption{
The same figure as Fig.~\ref{fig:specindexvsN1} for the model where an angular direction of a complex scalar field serves as the spectator field. 
}
\label{fig:specindexvsN2}
\end{figure*}

During inflation, the radial direction $\sigma$ obtains a Hubble-induced mass and rolls along the potential in Eq.~\eqref{eq:VH_corrected} while obtaining fluctuations. The angular direction $\theta$, owing to the approximate $U(1)$ symmetry, remains nearly massless and is frozen up to fluctuations produced during inflation.
After inflation, the radial direction is relaxed to the minimum of the potential with $\sigma \neq 0$ and its fluctuations are dampened.%
\footnote{
If the fluctuations of the radial direction are not dampened, then both the angular and radial fluctuations contribute to the
cosmic perturbations, resulting in a spectral index that falls between the values shown in Figs.~\ref{fig:specindexvsN1} and~\ref{fig:specindexvsN2}. 
}
For this to occur,  the potential of $\sigma$ after inflation should not be dominated by the potential in Eq.~\eqref{eq:VH_corrected}, where the mass of $\sigma$ around the minimum is small and the radial direction is not dampened. Instead, the dominant part of the potential should be the vacuum wine-bottle potential or one with an unsuppressed negative Hubble-induced mass term stabilized by a positive vacuum potential.
The mass of the angular direction is assumed to be still negligible and its fluctuations are frozen on super-horizon scales. Eventually, the mass of the angular direction arising from explicit $U(1)$ breaking becomes comparable to the Hubble scale and the fluctuations in the angular direction are converted into fluctuations in the energy density of $\Sigma$. This may occur as oscillations in the angular direction with fixed $\sigma$~\cite{Dimopoulos:2003az}, or as a spiral motion of $\Sigma$ in field space~\cite{McDonald:2003jk,Riotto:2008gs,Harigaya:2019uhf,Co:2022qpr} by the Affleck-Dine mechanism~\cite{Affleck:1984fy}.
The spiral motion, if the angular momentum is sufficiently large, is naturally long-lived because of the approximate $U(1)$ charge conservation~\cite{Co:2019wyp,Domcke:2022wpb} and naturally dominates the universe, making it a good curvaton candidate~\cite{Co:2022qpr}. Also, the $U(1)$ charge can be converted into baryon asymmetry without producing baryon isocurvature perturbations~\cite{Co:2022qpr}.

In the above setup, the angular direction is the spectator field, and hence its fluctuations are responsible for the cosmic perturbations, $\zeta \propto \delta \theta$.
The dynamics of the radial direction indirectly affects the spectrum of the fluctuations of the angular direction, which is given by
\begin{equation}
   \langle \delta \theta({\bm{k}}) \delta \theta({\bm{k'}}) \rangle = \frac{2\pi^2}{k^3} \delta^3({\bm{k}}-{\bm{k'}}) \frac{H^2}{\left(2\pi \sigma_*\left(k\right)\right)^2}.
\end{equation}
The spectrum of curvature perturbations is
\begin{equation}
    P_\zeta(k) \propto \frac{1}{\sigma_*(k)^2}.
\end{equation}
Then, the spectral index is given by
\begin{equation}
\label{eq:n_b3}
    n_s-1 = \frac{{\mathrm{d}} {\ln}P_\zeta(k)}{{\mathrm{d}}{\rm ln} k} = -2 \frac{{\rm d} {\ln}\sigma_*(k)}{{\mathrm{d}}{\ln} k} =  -2 \frac{{\rm d} {\ln} \sigma_*(N)}{{\mathrm{d}}{N} } = -2 \frac{{\mathrm{d}} {\ln} \{r_{\sigma}(N)\}}{{\mathrm{d}}N }. 
\end{equation}
Using Eq.~\eqref{eq:rsol_inf}, we obtain
\begin{equation}
\label{eq:n_b3_slow}
    n_s = 1 +\frac{4}{3}b \times  {\rm exp}\left( - \frac{2}{3}b N \right){\ln}(r_{\sigma,i}).
\end{equation}

\begin{figure*}
\centering
\includegraphics[width=250pt, trim={0.16cm 0.11cm 0.11cm 0.11cm}, clip] 
{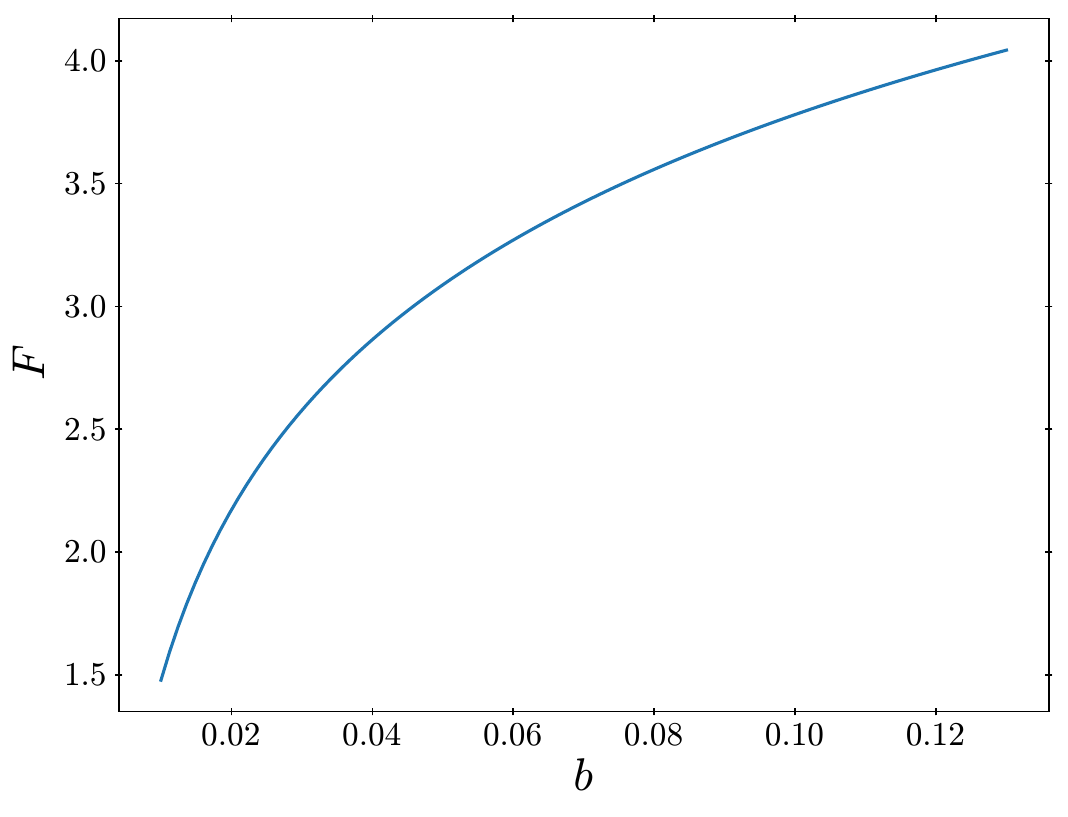}
\caption{The fine-tuning measure $F$ as a function of the model parameter $b$ in the model where an angular direction of a complex scalar field serves as the spectator field.}
\label{fig:ns_naturalness_axion}
\end{figure*}

\begin{figure*}
\centering
\includegraphics[width=270pt, trim={0.16cm 0.11cm 0.11cm 0.11cm}, clip] 
{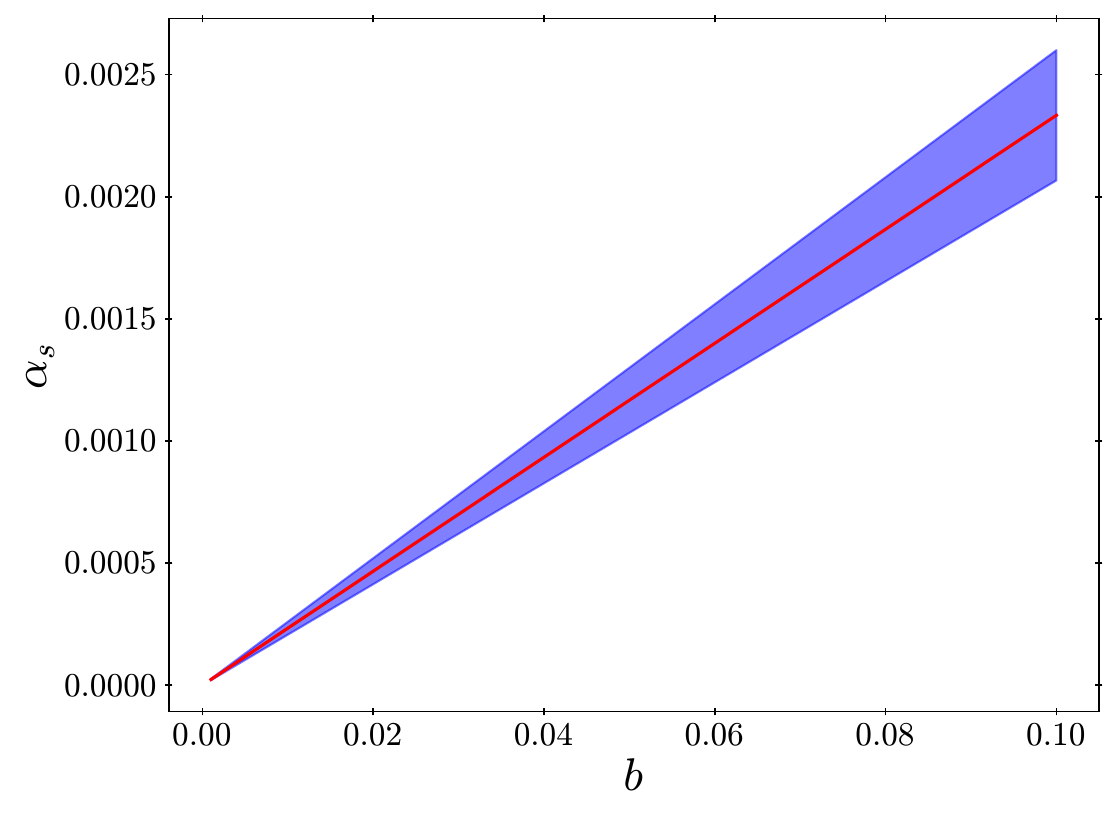}
\caption{
The red curve shows the running of the spectral index at $n_{s} = 0.965$ as a function of $b$ for the model where an angular direction of a complex scalar field serves as the spectator field. The blue-shaded region represents $n_{s}=0.965\pm0.004$, consistent with the $1\sigma$ \textit{Planck} constraints.
}
\label{fig:running_vs_b_axion}
\end{figure*}

The left panel of Fig.~\ref{fig:specindexvsN2} shows the spectral index as a function of the number of e-folds for different values of $b$, with $r_{\sigma,i}= 10^{-5}$. In this model, $n_{s}$ is smaller than $1$ since the radial direction stops evolving at the minimum of the potential.
The blue-shaded region shows the
 areas with $n_s=0.965 \pm 0.004$ as measured by \textit{Planck} at the $68\%$ confidence level, while the red-shaded region
indicates where $n_s=0.965 \pm 0.02$.
The right panel of Fig.~\ref{fig:specindexvsN2} shows
the number of e-folds corresponding to the shaded regions in the left panel.

Compared to the model presented in Sec.~\ref{sec:curvaton},
the number of e-folds between when the spectator field begins to roll from the origin and when
the CMB scale exits the horizon needs to be larger.
The red-shaded regions are wider than those in Fig.~\ref{fig:specindexvsN1}, indicating that the observed spectral index can be achieved more naturally in this setup.

We evaluate the naturalness of the observed spectral index as in Sec.~\ref{sec:curvaton}. Fig.~\ref{fig:ns_naturalness_axion} shows the fine-tuning measure $F$ as a function of the model parameter $b$.
The observed spectral index is naturally explained within this model even for $b$ as large as $0.1$, with the fine-tuning measure as low as $F=4$.

\subsection{Running of the spectral index}

The running of the spectral index is
\begin{equation}
    \alpha_s = -\frac{8}{9}b^2 \times {\rm exp}\left( - \frac{2}{3}b N \right){\rm ln}(r_{\sigma,i}) = \frac{2}{3}b (1-n_s), 
    \end{equation}
which is shown in Fig.~\ref{fig:running_vs_b_axion}.
The red curve shows $\alpha_s$ at $n_s=0.965$ and the blue-shaded region shows that for $n_s=0.965 \pm 0.004$.
An observable value of the running ($\alpha_s>10^{-3}$) can be produced for $b>0.04$ while naturally explaining the observed spectral index $(F\sim3)$.

\section{Summary}
\label{sec:conc}
In this work,
we propose a spectator field scenario where
quantum corrections to the scalar potential address
the eta problem while simultaneously explaining the slight deviation of cosmic perturbations from scale-invariance. We demonstrate that these quantum corrections induce an attractor behavior during inflation, driving the spectator field toward a region in field space where the curvature perturbations are nearly scale-invariant. If the initial condition of the spectator field is set at the origin of the field space, the resulting spectrum of curvature perturbations can be slightly red-tilted.
We further evaluate the naturalness of the observed spectral index by employing a fine-tuning measure and demonstrate that the observed spectral index can be obtained naturally within our model. 
Also, we show that an observable value of the running of the spectral index can be produced in a parameter region without fine-tuning. 

Spectator field models generically predict large non-Gaussianity of the cosmic perturbations.
We focus our analysis on a curvaton model with a quadratic vacuum potential and compute the primordial non-Gaussianity and running of the scalar spectral index. We derive a relationship between these two cosmological observables, providing an important theoretical prediction of the model that is testable with the next-generation cosmological probes.

We further demonstrate the versatility of our approach by applying it to a broader class of models where the angular direction of a complex scalar field with an approximate $U(1)$ symmetry serves as the spectator field.
Using the fine-tuning measure, we find that the observed spectral index can be obtained naturally also in this model by the attractor behavior of the radial direction of the complex scalar field, while possibly producing an observable value of the running.

\acknowledgments
We thank Aaron Pierce for commenting on the draft.
K.H.~is supported by the Department of Energy under Grant No.~DE-SC0025242, a Grant-in-Aid for Scientific Research from the Ministry of Education, Culture, Sports, Science, and Technology (MEXT), Japan (20H01895), and by World Premier International Research Center Initiative (WPI), MEXT, Japan (Kavli IPMU).


\appendix

\section{Inflation with large $\eta$}
\label{app:eta}

In this appendix, we demonstrate that the eta parameter of the inflaton may be $O(0.1-1)$ if the cosmic perturbations are not dominantly sourced by the inflaton fluctuations. 
Throughout, we work in units where the reduced Planck mass is unity.

\subsection{Model-independent analysis}

We begin with a model-independent analysis of the inflationary dynamics.
We consider an inflaton field $\phi$ with potential energy $\lambda \phi_e^4$, where $\phi_e^4$ denotes the field value at the end of inflation. The inflaton potential is 
\begin{equation}
\label{eq:Veff app}
    V(\phi) \simeq \lambda \phi_e^4 + \frac{3}{2}\kappa H_{\rm inf}^2 \phi^2 + \cdots,
\end{equation}
where the ellipses represent terms that trigger the end of inflation. In the limit where these terms are negligible, $\eta = \kappa$.
In hybrid inflation models, these are the interactions with the waterfall field, and in new inflation models, these are higher-order terms of the inflaton field. When $\kappa >0$, the inflaton field rolls towards smaller field values during inflation, which corresponds to hybrid inflation models.  
In contrast, when $\kappa <0$, the inflaton field rolls towards larger field values, which corresponds to new inflation models.
The Hubble scale during inflation is given by $H_{\rm inf}^2 \simeq \lambda \phi_e^4/3$, and the first slow-roll parameter is
\begin{equation}
\label{eq:first slow roll}
    \epsilon = \frac{\kappa^2}{2} \phi^2,
\end{equation}
which is much smaller than unity as long as $\phi < 1$.
The inflaton field value during inflation is
\begin{equation}
    \phi(N_e) \simeq \phi_e \times {\rm exp}\left(\frac{1}{2}(3 - \sqrt{9-12 \kappa})N_e\right),
\end{equation}
where the number of e-foldings $N_e$ is counted backward from the end of inflation. The curvature perturbation generated by the inflaton is
\begin{equation}
    {\cal P}_{\zeta,{\rm inf}} = \frac{1}{12\pi^2 \kappa^2} \frac{\lambda \phi_e^4}{\phi_*^2 }.
\end{equation}
The correction to the spectral index from the inflaton fluctuation is given by
\begin{equation}
    \Delta (n_s-1)= 2 \kappa \frac{ {\cal P}_{\zeta,{\rm inf}}}{ {\cal P}_{\zeta,{\rm obs}}}. 
\end{equation}
We require the magnitude of this correction to the spectral index from the inflaton to be smaller than $|\Delta(n_{s}-1)| \approx 0.04$, to avoid the need for fine-tuned cancellations between the contributions coming from the inflaton and the spectator field. We note that the resulting constraint on the parameter space is not highly sensitive to the precise value of this bound, due to the strong dependence of ${\cal P}_{\zeta,{\rm inf}}$ on $\phi_e$ and $\kappa$.  

It is well known that in order to explain the near homogeneity of the universe, the total number of e-foldings during inflation $N_{\rm tot}$ must be sufficiently large. Assuming instantaneous reheating after inflation, the bound on the total number of e-foldings is 
\begin{align}
    N_{\rm tot} > 57+ \frac{1}{2}{\rm ln} \left(\frac{H_{\rm inf}}{10^{10}~{\rm GeV}}\right)  \equiv N_{\rm min}.
\end{align}
We note that for lower reheating temperature, ${N_{\text{min}}}$ can take smaller values.

In supersymmetric theories, we also require that the soft supersymmetry-breaking mass of the inflaton at the vacuum be smaller than the Hubble scale during inflation to preserve the flatness of the inflaton potential.
The soft mass of the inflaton may be as small as the gravitino mass.
In gravity mediation, the gravitino mass must exceed the TeV scale to ensure that the Minimal Supersymmetric Standard Model (MSSM) particles are sufficiently heavy. In contrast, gauge mediation allows for much lighter gravitinos, with masses as small as $1$ eV~\cite{Ibe:2016kyg}. We therefore adopt $1$ eV as the lower bound on $H_{\text{inf}}$, while also using $H_{\text{inf}}= 1$ TeV as a benchmark value from gravity mediation.

\begin{figure*}
\centering
\includegraphics[width=430pt, trim={0.16cm 0.11cm 0.11cm 0.11cm}, clip] 
{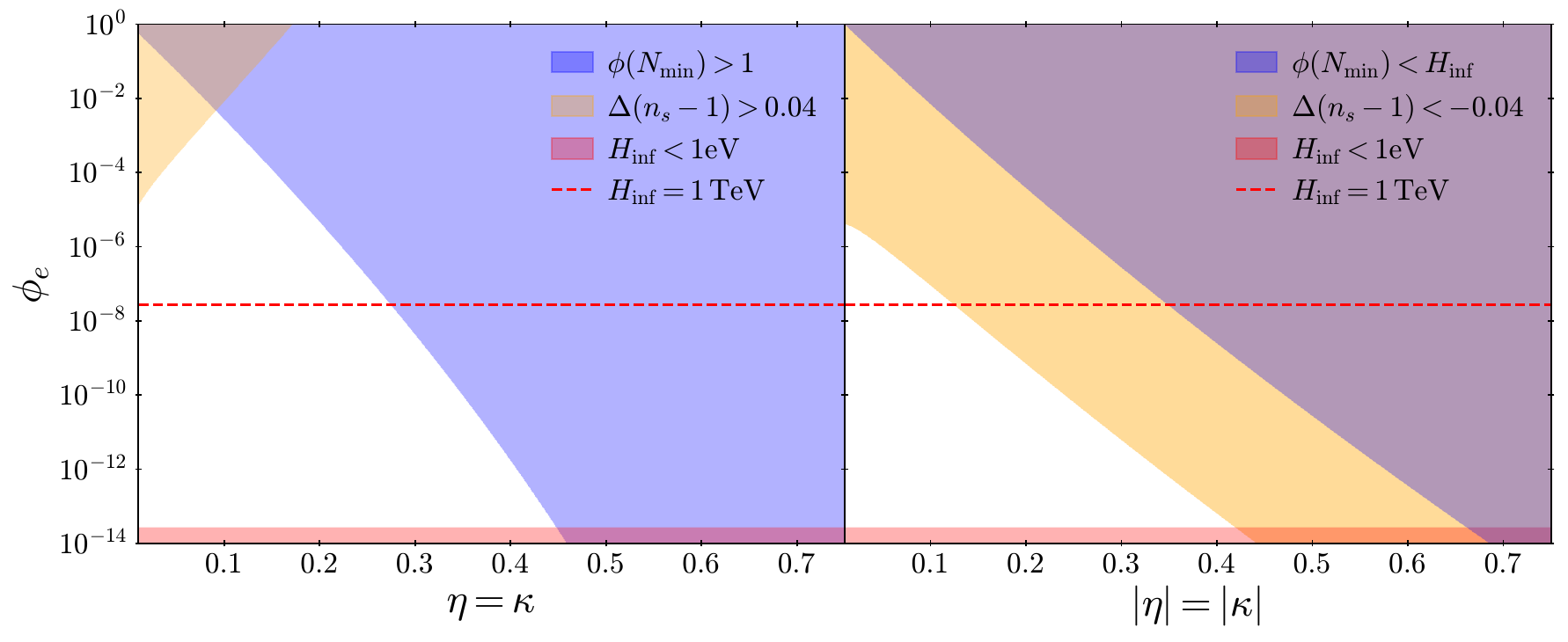}
\caption{The allowed parameter regions for hybrid (left) and new (right) inflation models. The blue-shaded region marks the parameter space where $\phi(N_{\rm min}) > 1$ (left) and $\phi(N_{\rm min}) < H_{\rm inf}$ (right). The orange-shaded regions represent $|\Delta(n_{s}-1)| > 0.04$, while the red-shaded regions mark where $H_{\rm inf} < 1$ eV in both panels. The red-dashed lines represent $H_{\rm inf} = 1$ TeV in both subplots.}
\label{fig:hybrid_new_inf_constraint}
\end{figure*}

The left panel of Fig.~\ref{fig:hybrid_new_inf_constraint} presents the allowed (unshaded) parameter regions for the hybrid inflation model ($\kappa >0$) with $\lambda=1$. The blue-shaded region corresponds to $\phi(N_{\rm min}) > 1$, for which higher-order terms in the inflaton potential may become significant. The orange-shaded region indicates where $\Delta(n_s-1) > 0.04$, corresponding to an unnatural
cancellation between the inflaton and spectator field contributions to explain the observed spectral index.
The red-shaded region shows where $H_{\rm inf}< 1$ eV, while the red-dashed line represents $H_{\rm inf} = 1$ TeV. The upper bound on $\kappa$ is determined by the blue and red-shaded regions.
For a given $H_{\rm inf}$ and $\lambda$,
the upper bound on $\kappa$ is approximately given by
\begin{equation}
    \kappa < \frac{0.36 + 0.005 \times{\rm ln}\lambda  - 0.010 \times {\rm ln}\left(\frac{H_{\rm inf}}{10^3~{\rm GeV}}\right)}{1 + 0.010 \times {\rm ln}\left(\frac{H_{\rm inf}}{10^3~{\rm GeV}}\right)}.
\end{equation}
Note that the maximal allowed value of $\kappa$ increases for larger $\lambda$. In the concrete model shown below, $\lambda$ is a coupling constant in the model and may be of order unity, allowing values of $\eta = \kappa = O(0.1-1)$.

Similarly, the right panel of Fig.~\ref{fig:hybrid_new_inf_constraint} shows the corresponding constraints for the new inflation model ($\kappa < 0$) with $\lambda=1$. 
The blue-shaded region denotes initial conditions with $\phi (N_{\rm min}) < H_{\rm inf}$, which are not physically realizable since quantum fluctuations typically induce field values of order $H_{\rm inf}$ or larger. The orange-shaded region indicates where $\Delta(n_{s}-1) < -0.04$. 
The upper bound on $|\kappa|$ is determined by the orange and red-shaded regions and is approximately given by
\begin{equation}
    |\kappa| < \frac{0.23 - 0.0056 \times{\rm ln} \left(\frac{\lambda}{10^{-6}}\right) - 0.011 \times {\rm ln}\left(\frac{H_{\rm inf}}{10^3~{\rm GeV}}\right) + 0.011 \times {\rm ln} \left(\frac{|\kappa|}{0.2}\right)}{1 + 0.01\times{\rm ln}\left(\frac{H_{\rm inf}}{10^3~{\rm GeV}}\right) }.
\end{equation}

Thus, values of $|\eta|= |\kappa|$ as large as $O(0.1-1)$ are permitted. Furthermore, smaller values of $\lambda$ expand the viable range of $|\eta|$.

\subsection{Hybrid inflation model}

The parameters $\lambda$ and $\phi_e$ can be related to the parameters in concrete models.
A hybrid inflation model can be realized in supersymmetric theory via the superpotential~\cite{Copeland:1994vg}
\begin{equation}
    W = y \Phi (\psi \bar{\psi} - v^2),
\end{equation}
where $\Phi$ is an inflaton chiral multiplet and $\psi$ and $\bar{\psi}$ are waterfall chiral multiplets. The parameters $\phi_e$ and $\lambda$ are given by $\phi_e = v$ and $\lambda = |y|^2$, respectively. Note that $\lambda < 1$ is required for the perturbativity of the model.
The eta parameter is determined by the Kahler potential of $\Phi$.

\subsection{New inflation model}

A new inflation model can be realized in supersymmetric theory via the superpotential~\cite{Kumekawa:1994gx}
\begin{align}
    W = v^2 \Phi - \frac{g}{n+1} \Phi^{n+1},
\end{align}
where $n$ is a positive integer larger than 2 and $g<1$ and $v^2 < 1$ are constants. The theory has $Z_{2nR}$ symmetry  under which $\Phi$ carries a charge $2$, and the inflaton may be initially trapped at the origin by a Hubble-induced mass or a thermal mass before the beginning of the last inflation. The inflaton potential from the superpotential is
\begin{equation}
    V \supset |v^2- g \Phi^{n}|^2.
\end{equation}
For a given $|\Phi|$, the potential is minimized for ${\rm Re}(\Phi)>0$ and ${\rm Im}(\Phi)=0$ (or the field values related to this by the $Z_{2nR}$ transformation). We thus take $\Phi = \phi/\sqrt{2}$ with $\phi>0$, for which the potential is
\begin{equation}
    V \supset v^4 - \frac{g}{2^{n/2-1}}\phi^n + \frac{g^2}{2^n}\phi^{2n}.
\end{equation}
For $\phi^n \ll v^2/g$, the third term is negligible. The contributions of the second term to the first and second slow-roll parameters are
\begin{align}
\label{eq:slow roll new}
    \epsilon \supset & \frac{n^2 g^2}{2^{n-1}} \frac{\phi^{2n-2}}{v^8}, \nonumber\\
    \eta \supset & - \frac{n(n-1)g}{2^{n/2-1}} \frac{\phi^{n-2}}{v^4},
\end{align}
respectively. Inflation ends when one of the slow-roll parameters exceeds unity. Since $\epsilon \propto \phi^{2n-2}$ while $\eta \propto \phi^{n-2}$, as long as $\phi <1$,  $\eta$ exceeds unity first. Thus, we have the following identification of parameters,
\begin{align}
    \phi_e = &  \frac{\sqrt{2}}{(n(n-1))^{1/(n-2)}} \left( \frac{v^4}{g} \right)^{\frac{1}{n-2}}, \nonumber \\
    \lambda = & \frac{1}{4} \left( n(n-1) g v^{n-6} \right)^{\frac{4}{n-2}}.
\end{align}
Small $\lambda$, which allows for large $|\eta|\simeq |\kappa|$, may be obtained by $v \ll 1$ for $n<6$ or by $g \ll 1$.
Unlike small $\eta$, smaller parameters in the superpotential are natural since their smallness may be protected by symmetry.

The $\epsilon$ and $\eta$ contributions from Eq.~\eqref{eq:Veff app} dominate over those from Eq.~\eqref{eq:slow roll new} until the very end of inflation. Therefore, the computation of the evolution of the inflaton field value and the curvature perturbation based on the potential in  Eq.~\eqref{eq:Veff app} is valid. As in the hybrid case, the eta parameter is dictated by the Kahler potential of $\Phi$.

\section{Hubble-induced mass after inflation}
\label{sec:HMD}

In this appendix, we derive the relation between the Hubble-induced masses of the spectator field $\sigma$ during and after inflation in supersymmetric one-field inflation models. In particular, we show that in these models the coefficient $b_{\rm m}$ matches $b$ and the Hubble-induced mass after inflation drives the spectator field to larger field values.

The inflaton field $\phi$ and the spectator field $\sigma$ are embedded into chiral multiplets $\Phi$ and $\Sigma$, respectively. Their Kahler and super potential are given by
\begin{equation}
    K = \left( 1 + f\left(\Sigma^\dag \Sigma\right)\right) \Phi^\dag \Phi + \Sigma^\dag \Sigma,~~ W  = W(\Phi)
\end{equation}
respectively, where $f$ is a function that encodes the tree and quantum level coupling between $\Phi$ and $\Sigma$.
The potential of $\phi$ and $\sigma$ is given by
\begin{align}
    V(\phi,\sigma) \simeq e^K \left( \frac{\partial^2 K}{\partial \Phi \partial \Phi^\dag}\right)^{-1} \left| \frac{\partial W}{\partial \Phi} \right|^2 \simeq   \left(1 + \frac{1}{2}\sigma^2-f\right) V(\phi),~~ V (\phi) = \left| \frac{\partial W}{\partial \Phi} \right|^2_{\Phi = \phi/\sqrt{2}},
\end{align}
where we assumed $|\phi|, |\sigma| \ll 1$ and kept the leading term in the inflaton potential and the Hubble-induced mass term of $\sigma$.
By matching this potential with Eq.~\eqref{eq:VH_corrected} using $V(\phi) = 3 H^2$ during inflation, we obtain
\begin{align}
    \frac{1}{2}\sigma^2-f = \frac{b}{3}\sigma^2 \left({\rm ln}\left(\frac{\sigma}{\sigma_0}\right)- \frac{1}{2}\right).
\end{align}
The kinetic term of the inflaton is given by
\begin{equation}
    \frac{1}{2}(1 + f) \partial \phi \partial\phi.
\end{equation}
After inflation, the Hubble-induced potential of $\sigma$ becomes
\begin{equation}
   \label{eq:VH_after}
    \left(\frac{1}{2}\sigma^2-f\right) V(\phi) - \frac{1}{2}f \dot{\phi}^2 = - \frac{3}{4} H^2 \sigma^2 + b H^2 \sigma^2 \left({\rm ln}\left(\frac{\sigma}{\sigma_0}\right)- \frac{1}{2}\right),
\end{equation}
where we used $V(\phi)\simeq \dot{\phi}^2/2 \simeq 3H^2/2$.
Comparing Eq.~\eqref{eq:VH_after} with Eq.~\eqref{eq:VH_MD},
one can see that $b_{\rm m}=b$ and the spectator field obtains an extra negative Hubble-induced mass with a coefficient $a_{\rm m}=3/2$, which drives the field to larger field values.

\section{Dynamics during matter domination}
\label{sec:Dynamics_MD}

In this appendix, we solve the equation of motion of $\sigma$ during the matter dominated era after inflation. The potential of $\sigma$ can be parameterized as
\begin{equation}
V = - \frac{1}{2}a_{\rm m} H^2 \sigma^2 + b_{\rm m} H^2 \sigma^2 \left({\ln}\left(\frac{\sigma}{\sigma_0}\right)- \frac{1}{2}\right).
\end{equation}
Then, the equation of motion of $\sigma$ is given by
\begin{equation}
\ddot{\sigma} + 3 H \dot{\sigma} - a_{\rm m} H^2 \sigma + 2 b_{\rm m} H^2 \sigma {\ln}\left(\frac{\sigma}{\sigma_0}\right)=0.
\end{equation}
Taking the number of e-folds as the time variable, the equation of motion becomes
\begin{equation}
\frac{{\mathrm{d}}^2 \sigma}{\mathrm{d}N^2} + \frac{3}{2} \frac{\mathrm{d} \sigma}{\mathrm{d}N} - a_{\rm m} \sigma + 2 b_{\rm m} \sigma {\ln}\left(\frac{\sigma}{\sigma_0}\right)=0.
\end{equation}
We denote the solution for $b_{\rm m}=0$ by $\bar{\sigma}$, which is given by
\begin{align}
\bar{\sigma}=& X(N)\sigma_{\rm end},\nonumber \\
X(N) =& \frac{3 + \sqrt{9 + 16 a_{\rm m}} }{2\sqrt{9 + 16 a_{\rm m}}} {\rm exp}\left\{\frac{- 3 + \sqrt{9 + 16 a_{\rm m}}}{4} (N-N_{\rm end})\right\} \nonumber \\
&+ \frac{-3 + \sqrt{9 + 16 a_{\rm m}} }{2\sqrt{9 + 16 a_{\rm m}}} {\rm exp}\left\{\frac{ -3 - \sqrt{9 + 16 a_{\rm m}}}{4} (N-N_{\rm end})\right\} \nonumber \\
  \simeq & \frac{3 + \sqrt{9 + 16 a_{\rm m}} }{2\sqrt{9 + 16 a_{\rm m}}} {\rm exp}\left\{\frac{- 3 + \sqrt{9 + 16 a_{\rm m}}}{4} (N-N_{\rm end})\right\},
  \label{eq:X_N}
\end{align}
where $\sigma(N_{\rm end}) \equiv \sigma_{\rm end}$, $\sigma'(N_{\rm end})\equiv0$, and we assumed $a_{\rm m} > -9/16$.
In the last equality, we neglect the second term in Eq.~\eqref{eq:X_N} since it decays faster than the other term.
Defining $s$ by $\sigma \equiv \bar{\sigma} s$, the equation of motion of $s$ is given by
\begin{equation}
\frac{\mathrm{d}^2 s}{\mathrm{d}N^2}+\left[ 2 \frac{\mathrm{d} {\ln}\bar{\sigma}}{\mathrm{d}N} + \frac{3}{2}\right] \frac{\mathrm{d}s}{\mathrm{d}N}+ 2 b_{\rm m} s\left[ {\ln} \left(\frac{\bar{\sigma}}{\sigma_0}\right)  + {\ln}s \right]=0.
\end{equation}
Under the slow-roll approximation for $s$ ($\frac{\mathrm{d}^2 s}{\mathrm{d}N^{2}} \simeq 0$), we find the evolution of $s$ is governed by
\begin{equation}
\frac{\sqrt{9 + 16 a_{\rm m}}}{2} \frac{\mathrm{d}{\ln}s}{\mathrm{d}N} + 2 b_{\rm m}  {\rm ln} \left(\frac{\bar{\sigma}}{\sigma_0}\right) + 2 b_{\rm m} {\ln}s  =0,
\end{equation}
which can be solved as
\begin{equation}
{\ln} \{s(N)\} = \left(-1 +{\rm exp} \left\{ -\frac{4 b_{\rm m}}{\sqrt{9 + 16 a_{\rm m}}} (N - N_{\rm end}) \right\}\right){\ln}\left(\frac{\sigma_{\rm end}}{\sigma_0}\right) + C,
\end{equation}
where $C$ is a constant that is independent of $\sigma_{\rm end}$. The solution for $\sigma$ is then
\begin{equation}
{\ln}\left(\frac{\sigma}{\sigma_0}\right) =  {\rm exp} \left\{ -\frac{4 b_{\rm m}}{\sqrt{9 + 16 a_{\rm m}}} (N -N_{\rm end}) \right\}{\ln}\left(\frac{\sigma_{\rm end}}{\sigma_0}\right) + C',
\end{equation}
where $C'$ is another constant also independent of $\sigma_{\rm end}$.
Using this formula, we may derive the prediction on the local non-Gaussianity, as shown in Sec.~\ref{sec:curvaton}.


\bibliographystyle{jhep}
\bibliography{mybib} 
\end{document}